# Acoustic Coherent Perfect Absorbers as Sensitive Null Detectors

Chong Meng, Xiaonan Zhang, Suet To Tang, Min Yang, and Zhiyu Yang*

Department of Physics, the Hong Kong University of Science and Technology

Clearwater Bay, Kowloon

Hong Kong, China

**Abstract**

   We report the experimental realization of acoustic coherent perfect absorption (CPA) of four symmetric scatterers of very different structures. The only conditions necessary for these scatterers to exhibit CPA are that both the reflection and transmission amplitudes of the scatterers are 0.5 under one incident wave, and there are two collinear and counter-propagating incident waves with appropriate relative amplitude and phase. Nearly 1000 times in the modulation of output power has been demonstrated by changing the relative phase of the incident waves over 180°. We further demonstrate that these scatterers are sensitive devices to detect the small differences between two nearly equal incident waves. A 27 % change in the strength of the scattering wave has been demonstrated for every degree of phase deviation from the optimum condition between the incident waves.

Wheatstone bridge is a classic electric circuit that can detect small differences between the two circuit arms in terms of tiny electric current via a galvanometer across the bridge. Its analogy in wave physics could be utilized to detect the small differences between two coherent waves (arms) that have the same frequency and highly correlated relative amplitude and phase. The wave devices that can serve such function are made possible by the coherent perfect absorption (CPA) scatterers, where a small difference in the incident waves will lead to the generation of detectable outgoing waves that are otherwise absent when the two incident waves are perfectly matched in amplitude and phase [1]. The CPA process was originally proposed as the time-reversed process of optical lasing at threshold. When a scatterer is illuminated by two collinear and counter-propagating incident electromagnetic waves with controlled relative phase and amplitude, which act as the time reversed output waves of a lasing mode, perfect absorption of the incident waves is predicted. The theoretical concept [1] was first experimentally realized by using a silicon cavity [2]. Later, deep sub-wavelength scale absorbers [3] and all-dielectric metasurface absorbers [4] were predicted and experimentally verified. The CPA frequency of the metasurface could be tuned from 8.56 to 13.47 GHz by solely adjusting the thickness of the metasurface [4]. Owing to the stringent conditions for CPA that a small mismatch in amplitude and/or phase in the two incident waves can be significantly manifested in the outgoing wave amplitude, such devices are potential candidates as sensitive optical bridge detectors for a variety of potential applications, or as coherent control of loss enhancement and suppression devices. Indeed, such control has been demonstrated on the optical absorption in an array of metallic nano-antennas covered by a thin luminescent layer [5].

The acoustic counterpart of optical CPA scatterers can be analyzed by the generic one dimensional two-port input/output system with a scatterer at the center. The scatterer can be described by the transfer matrix that relates the amplitudes of the incident waves $p_i^L$ and $p_i^R$ on the left and the right side of the scatterer to the outgoing ones $p_o^L$ and $p_o^R$ in the form $\begin{pmatrix} p_o^L \\ p_o^R \end{pmatrix} = \begin{pmatrix} t & r_L \\ r_R & t \end{pmatrix} \begin{pmatrix} p_i^R \\ p_i^L \end{pmatrix}$, where $t$ is the transmission of the scatterer, and $r_L$ ($r_R$) is the reflection at the left (right) surface of the scatterer under one wave incidence. Based on the simple acoustic CPA scheme [6], there are two possible types of CPA scatterers with mirror structural symmetry. For monopoles the two incident waves must be in phase, and the one-wave reflection $r$ and the transmission $t$ of the scatterer follows the relation $r = -t$. For dipoles the two incident waves must be out of phase, and the one-wave reflection follows the relation $r = t$. The one-wave transmission and reflection of a scatterer is independent of the amplitude of the incident wave in the linear acoustics regime, so they can be regarded as the characteristics of the scatterer. Combined with the constraint on $r$ and $t$ due to monopolar or dipolar symmetry [6], the magnitudes of $r$ and $t$ must be 0.5 for CPA, which leads to maximum absorption ($A_1 = 1 - |t|^2 - |r|^2$) of 0.5 under one incident wave. This 'yard stick' was used as the indicator of reaching CPA in the one-wave nonlinear absorption experiments on Helmholtz resonator (HR) [7]. Absorption of a scatterer reaching 0.5 under one incident wave is a necessary condition for the scatterer to achieve CPA, and the realization of it is an important step in the right direction. However, we do not regard it as full realization of CPA because no manipulation of absorption due to the coherent interplay between two incident waves is demonstrated. For the same reason, the uni-direction total absorption [8 – 10], even though being an extreme case of CPA with transmission and one surface reflection (say $r_L$) being zero, while the other surface being a perfect reflector, cannot be regarded as CPA.

In this paper, we report four types of scatterers that can lead to CPA under the conditions precisely predicted by the simple scheme [6], with their monopolar/dipolar nature confirmed by the experimentally measured one-wave transmission and reflection. We further demonstrate the operation of an acoustic bridge with the two incident waves as the two arms and a CPA scatterer as the null detector for the measurements of the amplitude and phase mismatch between the two arms.

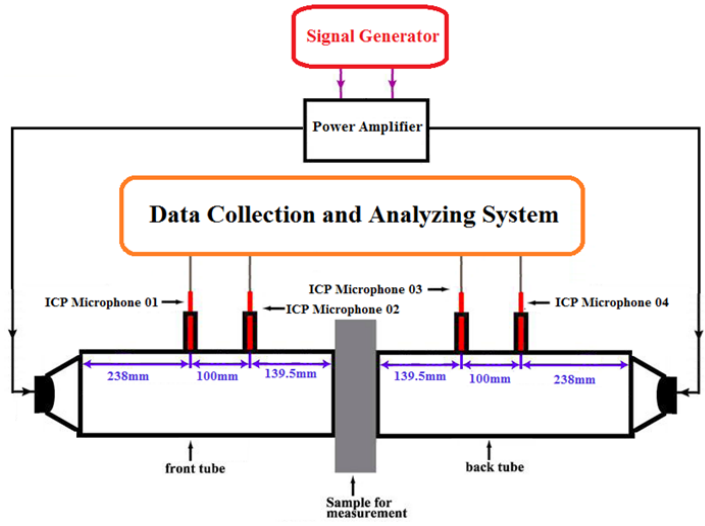

Figure 1. Schematics of the two-port acoustic setup for CPA.

The experimental setup for the measurements of the reflection and transmission under one incident wave is the same as we have previously reported [8]. For CPA experiments, we used a home-built symmetric impedance tube system shown in Fig. 1. The impedance tube along with the loudspeaker and microphones on the left side of the sample were matched as closely as possible to their counterparts on the right side of the sample. A piece of sponge was placed in front of each loudspeaker to reduce the reflection and tube resonance. The amplitudes of the waves emitted from the two loudspeakers along with their relative phase could be adjusted to any value. As has been well established in standard test methods, from the signals of the four strategically located microphones one can obtain the relative amplitudes between the four above defined waves $p_i^L$, $p_i^R$, $p_o^L$, and $p_o^R$ in the impedance tubes. The details are presented in Ref. 11. For simplicity, we take the amplitude of $p_i^L$ to be 1 in the remaining part of this paper. The absorption under such scheme is then $A_2 = 1 - \frac{|p_o^L|^2 + |p_o^R|^2}{|p_i^L|^2 + |p_i^R|^2}$, which we refer to as the CPA coefficient. The occurrence of CPA is marked by the diminishing outgoing wave amplitudes $p_o^L$ and $p_o^R$, and the CPA coefficient $A_2$ approaching 1. The absorption would depend on the relative phase and the amplitude ratio between the two waves emitted by the two loudspeakers, which is a unique characteristics of coherent absorption that can be utilized as the null detector in an acoustic bridge.

The four samples for the CPA experiments, although very different in structures, are simple monopoles or dipoles consisting of decorated membrane resonators (DMR's) [12]. For simple monopoles and dipoles, the geometric constraints dictate that their absorption cannot exceed 0.5 [10]. At maximum absorption, the transmission and the reflection amplitudes are both equal to 0.5, fulfilling the CPA condition. The DMR's offer the flexibility to adjust the transmission and the reflection of these samples over a wide range by the design parameters, such as the size of the membrane of the DMR, the size and the mass of the decorating platelet, the way they are mounted (on the sidewall or intersecting the waveguide), and in some cases the volume of the sealed air cavity. These parameters determine the surface response function [13] and the waveguide acoustic impedance, while the curvature energy at the edges of the membrane provides the dissipation [14]. In general, the DMR's with larger membranes tend to have larger resonant strength, which causes larger reflection and smaller transmission at resonance when they are mounted on the sidewall of the waveguide. The opposite is true when they are placed intersecting the waveguide. We can further fine-tune the transmission and reflection by small change of the mass of the decorating platelet to reach the optimum values for CPA.

Numerical simulations (COMSOL MultiPhysics) were carried out to verify the underline mechanism for CPA in these devices. Actual device structures parameters were used in the simulations. The mass density, Poisson's ratio, Young's modulus, and the pre-stress of the membrane were 1300 $kg/m^3$, 0.47, $1.2 \times 10^5$, and $0.3 MPa$, respectively. The dissipation is introduced in the form of an imaginary part in the tension that is about 1 % of the real part.

The first monopole sample for CPA is a hybrid membrane resonator (HMR) [9] mounted on the sidewall of a square waveguide that has the same cross section as the impedance tubes, as shown in the insert of Fig. 2(a). The HMR consists of a DMR backed by a sealed cavity. The cylindrical cavity is 36 mm in radius and 58 mm in depth. The front surface is sealed by a 0.2 mm thick stretched rubber membrane decorated at the center with a platelet of 3.6 mm in radius and 119 mg in mass. The one-wave transmission and reflection spectra of the sample are shown in Fig. 2(b). For clarity, the phase spectrum of the reflection has been down shifted by 180°. The points are experimental amplitude and phase spectra, and the solid and the dashed curves are numerical results. They are seen to agree well. The relative amplitudes of the transmission and reflection at resonance can be adjusted by changing the mass of the platelet, which also leads to trivial shift in the resonant frequency. At 210.95 Hz, the resonant transmission is 0.47 and the reflection is 0.53. The phase difference is 181.2°, which manifests the monopolar nature of the sample. The sample is then ready for monopolar CPA experiments. The phase difference obtained by simulations at the same frequency is 179.6°, and both the amplitudes are at the perfect value of 0.5.

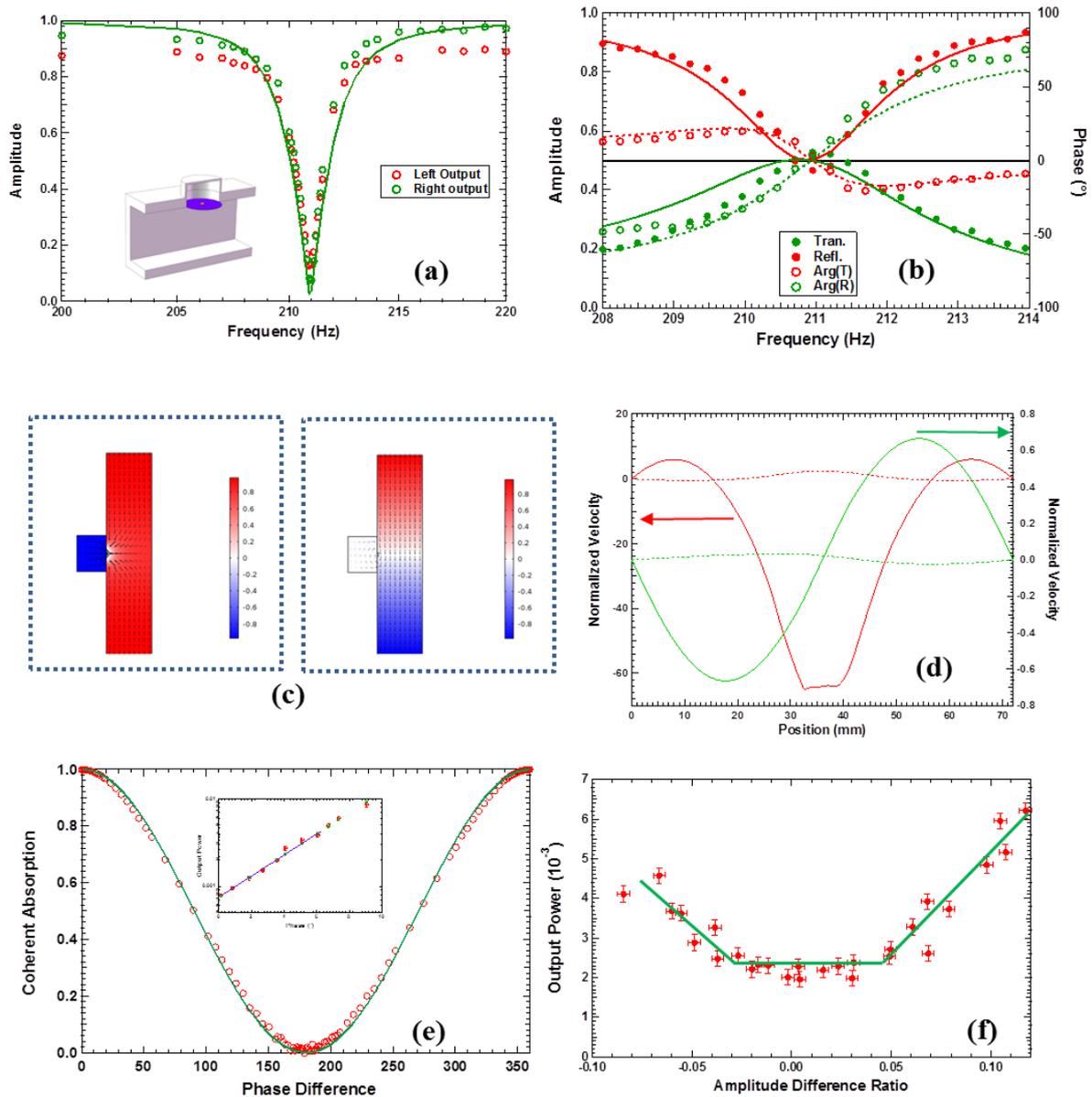

Figure 2. CPA of a side mounted monopole scatterer. (a) The experimental amplitudes of the outgoing waves as a function of frequency (points) under CPA conditions for monopoles. The solid curve is from simulations. The insert is a schematic of the sample. (b) The one-wave reflection and transmission spectra (points) and numerical simulation results for the amplitude (solid curves) and the phase (dashed curves). (c) The velocity (arrows) and pressure (color coded) fields when the two input waves are in-phase (left) and 180° off-phase (right). (d) The vibration profiles of the membrane at CPA (red curves, left axis) and off CPA (green curves, right axis). The solid curves are for the real part of the velocity and the dashed ones are for the imaginary part. (e) The CPA coefficient as a function of the relative phase between the two incoming waves at optimum matching amplitudes. The insert depicts the intensity of the outgoing waves as a function of small phase deviation near the optimum coherent absorption conditions. (f) The intensity of the outgoing waves as a function of amplitude mismatch of the two incoming waves. The solid lines are for guiding the eyes only.

The measured amplitudes of the outgoing waves when both loudspeakers were emitting acoustic waves of proper relative amplitude and relative phase for CPA are shown as points in Fig. 2(a). The relative phase of the two speakers was set to the optimum value of 0.2°, which is very close to the perfect in-phase condition of 0° for monopolar CPA. The amplitudes of the two incident waves were controlled to be within 2 % in relative deviation. CPA occurred at exactly the same frequency of 210.95 Hz with $p_i^R = 0.999$, $p_o^L = 0.032$, and $p_o^R = 0.034$. The ratio of input energy over the output energy [6] is nearly 1000. Numerical simulation results (curves in Fig. 2(a)) agree well with the experimental ones.

The air velocity field obtained by numerical simulations under optimum CPA conditions is shown in the left portion of Fig. 2(c). Near the platelet the air velocity components along the waveguide axis are opposite in sign, implying monopolar mode. The pressure field inside the cavity is 180° out of phase to the one on the opposite side of the membrane. The corresponding velocity profile (red curves) of the membrane across its central line parallel to the waveguide axis is shown in Fig. 2(d). The profile clearly resembles the hybrid resonance of the HMR [9] coupled with the air in the waveguide. The real part of the velocity of the platelet at CPA is over 60 times the far field air velocity of the incident waves, which reveals the dissipation mechanism of CPA. The pressure and the velocity fields when the input waves are off-phase by 180° are plotted in the right portion of Fig. 2(c). The overall field is of dipolar symmetry near the scatterer, with a node at the position of the platelet. The waves are hardly disturbed by the scatterer. As shown by the green curves in Fig. 2(d), there is only small movement of the platelet, and the vibration profile of the DMR is anti-symmetric following the characteristics of the out-of-phase dipolar incoming waves. The maximum amplitude is only 0.6 times of the far field incoming waves, leading to near zero absorption as expected.

When a relative phase $\phi$ is introduced between the two incoming waves, the outgoing waves are given by $\begin{pmatrix} p_o^L \\ p_o^R \end{pmatrix} = \begin{pmatrix} 1/2 & -1/2 \\ -1/2 & 1/2 \end{pmatrix} \begin{pmatrix} 1 \\ e^{i\phi} \end{pmatrix}$. The amplitude of the outgoing waves are then proportional to $\sin(\phi/2)$, and the CPA coefficient is of the form $A_2 = \cos^2(\phi/2)$. This is confirmed by the experimental CPA coefficient at different relative phase values, as shown in Fig. 2(e). It is seen that the experimental data (circles) agree well with the prediction of the simple theory (curve). Furthermore, the CPA coefficient is very sensitive to phase mismatch, so it can be used as a detector in an acoustic bridge setting with the two incoming waves as the two arms. The insert of Fig. 2(e) depicts the magnified section near the optimum phase for minimum output. The points are experimental data and the line is a fit to a single exponential function. It is seen that within 10° of mismatch the output power increases exponentially at a rate of 0.28 times per degree of phase mismatch. This indicates that the phase coherence of the two incident waves is very stable, with estimated fluctuation below 0.2°.

We then tested the dependence of CPA on the amplitude mismatch between the two incident waves. The experimental results, which are shown in Fig. 2(f), show that the amplitudes of the outgoing waves remain unchanged when the amplitude mismatch between the incident waves is less than 5 %. Above this 'threshed' the output increases linearly with the mismatch. This may suggest that the residual output obtained at the optimum experimental CPA conditions is partly due to the fluctuation of the amplitudes of the sound waves emitted by the two speakers, which we estimated as of the order of 5 %. The CPA scatterer could therefore detect the stability of both the phase and the amplitude of the sound sources.

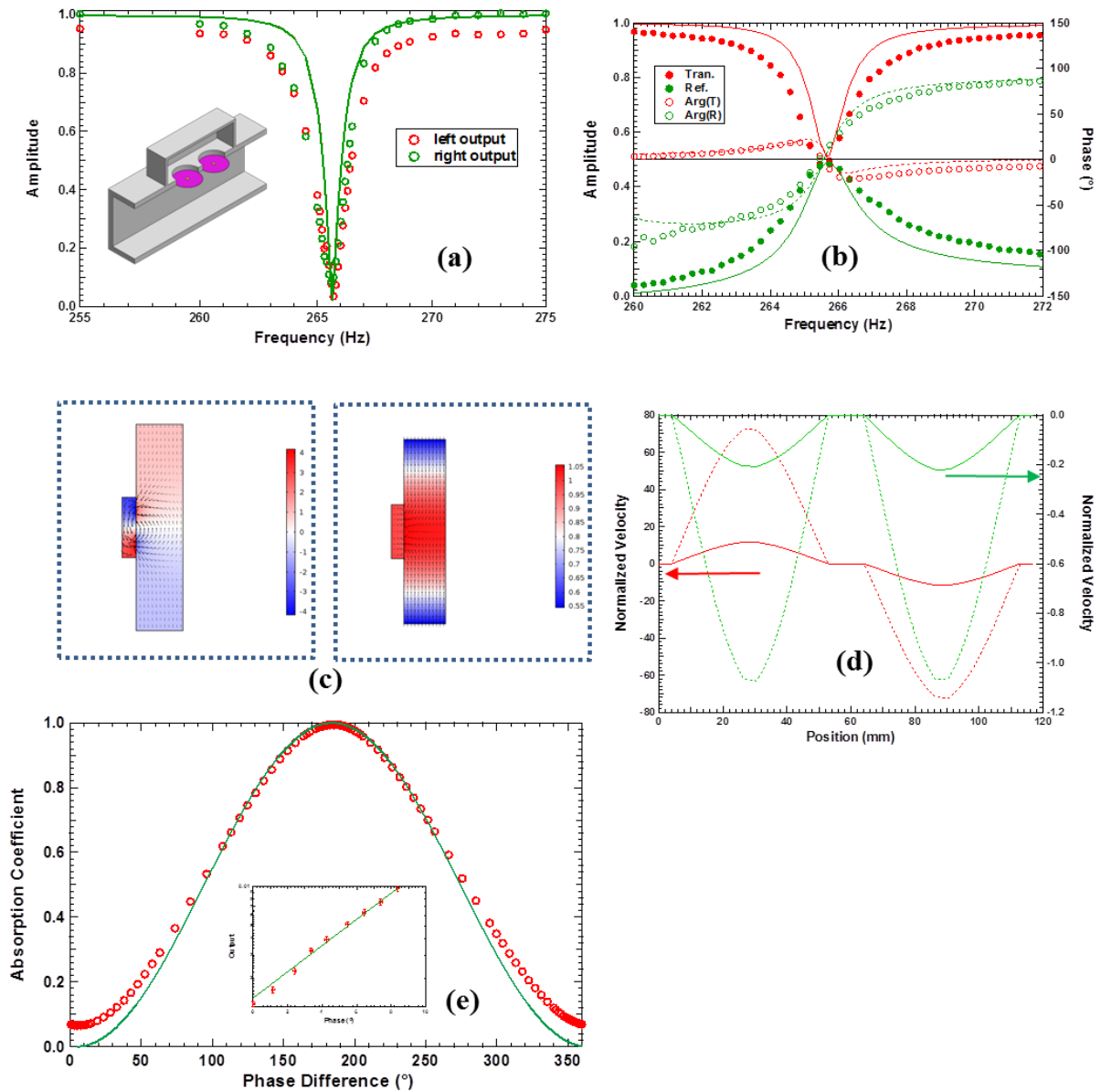

Figure 3. CPA of a dipole scatterer. (a) The experimental amplitudes of the outgoing waves as a function of frequency (points) under CPA conditions for dipoles. The solid curve is from simulations. The insert is a schematic of the sample. (b) The one-wave reflection and transmission spectra (points) and the numerical simulation results for the amplitude (solid curves) and the phase (dashed curves). (c) The velocity (arrows) and the pressure (color coded) fields when the two input waves are 180° off-phase (left) and in-phase (right). (d) The vibration profiles of the membrane at CPA (red curves, left axis) and off CPA (green curves, right axis). The solid curves are for the real part of the velocity and the dashed ones are for the imaginary part. (e) The CPA coefficient as a function of the relative phase between the two incoming waves at optimum matching amplitudes. The insert depicts the amplitude of the outgoing waves as a function of phase deviation near the optimum coherent absorption conditions.

Next, we built the dipole that also contains clear channel for airflow, as shown in the insert of Fig. 3(a). The front surface of the rectangular cavity is 117 mm × 56 mm, and the depth is 27 mm. Two membranes 24.5 mm in radius seal the holes on the front surface at 60 mm in distance from center to center. A platelet is attached to the center of each membrane. Each platelet is 6 mg in mass and 4 mm in diameter. The one-wave transmission and reflection spectra are shown in Fig. 3(b). At 265.45 Hz, both the transmission and the reflection are 0.48, with a phase difference of 3.2°. Simulation results are $r = 0.49$, $t = 0.52$, and the phase difference is 1.5°. Apart from the line width which is about half of that of the experimental spectra, the simulation results reproduce the essential features of the experimental spectra, including the phase spectra, and are in overall agreement with the experimental ones.

The amplitudes of the outgoing waves under CPA conditions are shown in Fig. 3(a), when the relative phase of the two speakers was set to the optimum value of 180.3°, which is very close to the perfect off-phase condition of 180° for dipolar CPA. CPA occurred at 265.70 Hz, with $p_i^R = 0.990$, $p_o^L = 0.0775$, and $p_o^R = 0.0345$. The CPA coefficient is 99.6 %. The ratio of input energy over the output energy is 278. The simulation results are seen to agree well with the experimental ones except for the somewhat narrower line width, which is also present in the one-wave transmission and reflection spectra.

The pressure and the air velocity fields from numerical simulations are shown in Fig. 3(c). The left part of the figure depicts the fields under dipolar CPA conditions where the two input waves are 180° out of phase. It is seen that near the two platelets the air velocity components along the waveguide axis are in the same direction, implying dipolar mode. Inside the cavity, the pressure field is clearly dipolar, i. e., the pressure in the left half of the cavity is 180° out of phase to the pressure in the other half. Each membrane vibrates in unison with its attached platelet. The two membranes are vibrating 180° out of phase to each other. The maximum amplitude is over 70 times of the far field incoming waves, which is a clear indication of resonant dissipation. When the input waves are perfectly in phase, the platelets are almost motionless (the right figure of Fig. 3(c)), and the scatterer is completely inert, which leads to expected zero absorption.

The dependence of the CPA coefficient on the phase difference between the two incoming waves is similar to the simple theoretical model for the monopole case. It is seen in Fig. 3(e) that the experimental data agree well with the theoretical prediction. The sample could also be used as a phase-sensitive detector, as indicated by the insert in Fig. 3(e) which is a magnified portion near the optimum phase difference. The points are experimental data and the curve is a fit to an exponential function. The change of the output power is about 0.23 times per degree of phase mismatch, up to about 10°. The rate of increase is a bit smaller than the first case, due to the somewhat larger residual output at optimum CPA conditions.

The above two cases are 'transparent' scatterers in that their resonance causes the transmission to reduce from nearly 1 to 0.5 while increasing the reflection from nearly 0 to 0.5. In the following, we briefly present the CPA results of two samples that are 'opaque', in that at resonance their transmission increases from low to 0.5, while their reflection decreases from nearly 1 to 0.5. The procedure to realize CPA for the two samples is the same as the previous two. The one-wave transmission and reflection spectra were measured first to ensure near 0.5/0.5 matching, while their phase difference dictated the parity of the two incident waves for CPA.

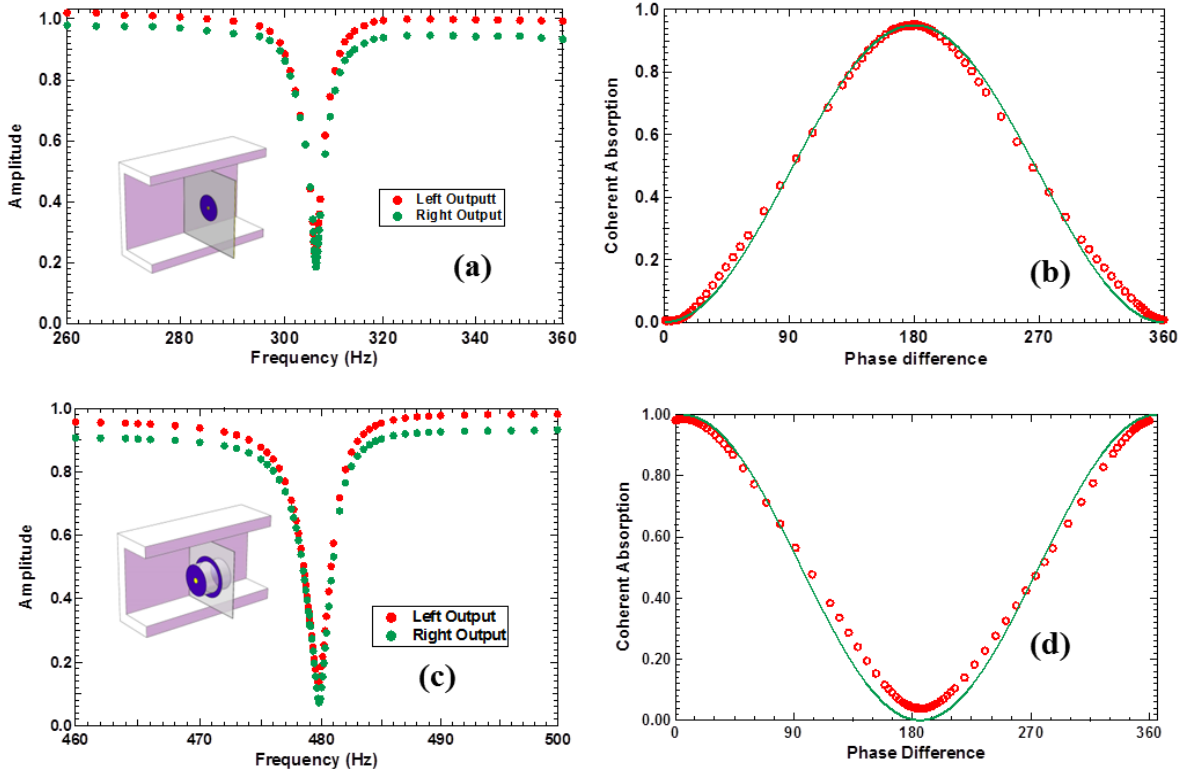

Figure 4. (a) The CPA of an 'opaque' dipole. The insert shows its structure. (b) The CPA coefficient of the 'opaque' dipole as a function of phase difference between the two incident waves at optimum amplitude ratio. (c) The CPA of an 'opaque' monopole. The insert shows its structure. (d) The CPA coefficient of the 'opaque' monopole as a function of phase difference between the two incident waves at optimum amplitude ratio.

The third symmetric CPA scatterer is just a simple DMR. Unlike the others, this sample does not contain any cavity. A circular rubber membrane of radius 19 mm was uniformly stretched on a solid frame. Two identical platelets, each of 3.6 mm in radius and 60 mg in mass, were symmetrically bonded to the center on each side of the membrane. Its sketch is shown in the insert of Fig. 4(a), together with the amplitudes of the outgoing waves under CPA conditions when the relative phase of the two speakers was set to the optimum value of 180.3°, which is very close to the perfect phase condition of 180° for dipolar CPA. CPA occurred at exactly the same frequency of 158.60 Hz where the transmission and the reflection are both close to 0.5, with $p_i^R = 0.99$, $p_o^L = 0.18$, and $p_o^R = 0.10$. The CPA coefficient is 95.8 %. The ratio of input energy over the output energy is over 47. The dependence of the CPA coefficient on the phase difference between the two incoming waves is the same as that for the previous dipole case. It is seen in Fig. 4(b) that the experimental data agree well with the theoretical prediction.

The fourth sample is a monopole similar to the one reported earlier [8]. Its sketch is shown in the insert of Fig. 4(c). A 0.2 mm thick ring-shaped stretched membrane supports a hollow rigid cylinder in the center. The radius of the clamped outer rim of the membrane is 27.4 mm and its inner rim is fixed on the tube with an outer radius of 21.4 mm. The hollow cylinder is sealed by two identical DMR's, one on each end. The radius of the two membranes is 20 mm, and the central platelets is 3.8 mm in radius and 70 mg in mass. The cylinder is 40 mm in length and 7.76 g in mass. The sample has a number of dipolar and monopolar resonances below 1500 Hz. The lowest monopole is at 248.5 Hz with transmission and reflection close to 0.5. The amplitudes of the outgoing waves under CPA conditions are shown in Fig. 4(c), when the relative phase of the two speakers was set to the optimum value of 5.8°, which deviates by a small margin from the perfect in-phase condition of 0° for pure monopolar CPA. This is because a nearby dipole resonance also contributed in the vibration. CPA occurred at 248.5 Hz with $p_i^R = 1.02$, $p_o^L = 0.14$, and $p_o^R = 0.07$. The CPA coefficient is 98.8 %. The ratio of the input energy over the output energy is 83. Due to the fact that the sample is not a pure monopole, the amplitude of the incoming wave from the right side was adjusted by up to 10 % relative to that from the left side in order to produce the nearly identical line shapes of the two outgoing waves near the CPA frequency. The dependence of the CPA coefficient on the phase difference between the two incoming waves is shown in Fig. 4(d). It is seen that apart from an offset angle of 5.8° and a small absorption background unrelated to CPA, the experimental data agree well with the theoretical prediction.

The residual outputs of the last two samples are several times larger than the first two samples at the optimum CPA conditions. They are therefore not as effective for acoustic bridges. It seems that the configurations of these two samples are more sensitive to the imperfections of the sample structures, such as the position of the platelet that may be off the center by a small margin, and the non-uniformity of the tension distribution in the membrane. Further studies are being conducted to resolve these issues.

A CPA scatterer is essentially a null detector with outgoing waves as its output without using a power supply for itself. A dipolar CPA scatterer functions the same manner as a power detector for two interfering input waves, i. e., the output reaches minimum when the two incoming waves are 180° off phase and with the same amplitude. The output reaches maximum when the two waves are in-phase, and the CPA is completely turned off. A monopolar CPA scatterer, however, is exactly the opposite. Minimum output occurs when the two incoming waves are exactly of the same amplitude and perfectly in-phase, which should lead to maximum signal of the power detector. Therefore, a monopolar CPA scatterer operates in complementary to a power detector. It is at its most sensitive detection range as a null detector when a power detector is at its least sensitive range with signal output near maximum. Furthermore, if one can incorporate additional components into a CPA scatterer to convert the vibration energy of the platelet into electric energy, such as those presented in Ref. 9, the CPA could then function as a dual-output detector, with the two output signals being exactly complementary to each other. At maximum CPA, the outgoing wave energy flux is minimum, while the electric signal from the vibrating platelet is maximum, and vice versa.

The two forms of outputs could be used alternatively for maximum sensitivity for the two extreme cases of optimum CPA and completely off-CPA.

In summary, we have experimentally demonstrated four types of scatterers for acoustic CPA, using DMR as the basic building blocks. The use of DMR's allowed us to design the samples with the transmission, reflection, and absorption properties suitable for CPA conditions. Nearly 1000 times in the modulation of output power has been demonstrated by changing the relative phase of incident waves over 180°. We have further demonstrated that, in an acoustic bridge configuration, these scatterers are sensitive devices to detect the small differences between two nearly equal incident waves. For the best device, a 27 % change in the strength of the output wave has been demonstrated for every degree of deviation from the optimum condition. The side-mount configuration of the first two CPA scatterers offers the possibility to mount multiple scatterers on the four sidewalls to realize multiple-frequency CPA by a single composite scatterer. The experimental realization of acoustic CPA with a variety of scatterers of different structures has opened the door to the possibility of further coherent manipulations of classical waves, including CPA by multiple scatterers, each is capable of CPA when used alone.

Acknowledgement – We sincerely thank P. Sheng for invaluable suggestions. This work was supported by AoE/P-02/12 from the Research Grant Council of the Hong Kong SAR government.